\documentclass[twocolumn,aps,prl,superscriptaddress,showpacs,floatfix]{revtex4}
\usepackage{amsmath,bm}
\usepackage{graphicx}

\begin{document}

\title{Parton coalescence and the antiproton/pion anomaly at RHIC}
\author{V. Greco}
\affiliation{Cyclotron Institute and Physics Department, Texas A\&M 
University, College Station, Texas 77843-3366, USA}
\author{C. M. Ko}
\affiliation{Cyclotron Institute and Physics Department, Texas A\&M 
University, College Station, Texas 77843-3366, USA}
\author{P. L\'evai}
\affiliation{Cyclotron Institute and Physics Department, Texas A\&M 
University, College Station, Texas 77843-3366, USA}
\affiliation{KFKI Research Institute for Particle and Nuclear Physics,
P.O. Box 49, Budapest 1525, Hungary}

\date{\today}

\begin{abstract}
Coalescence of minijet partons with partons from the quark-gluon 
plasma formed in relativistic heavy ion collisions is suggested as the 
mechanism for production of hadrons with intermediate transverse momentum. 
The resulting enhanced antiproton and pion yields at intermediate 
transverse momenta gives a plausible explanation for the observed 
large antiproton to pion ratio. With further increasing momentum, 
the ratio is predicted to decrease and approach the small value given 
by independent fragmentations of minijet partons after their energy 
loss in the quark-gluon plasma.
\end{abstract}

\maketitle

Heavy ion collisions at the Relativistic Heavy Ion Collider (RHIC) provides
the possibility of creating in the laboratory a plasma of deconfined quarks
and gluons. One of the signatures for the quark-gluon plasma (QGP) is
suppression of jet production \cite{wang}. Experimental data on high 
transverse momentum hadrons \cite{phenix,star}, which are dominantly 
produced from minijet partons originated from initial hard 
processes between colliding nucleons, have indeed shown a large suppression 
compared to what is expected from the superposition of primary binary 
nucleon-nucleon collisions. The amount of energy loss of minijet partons, 
particularly gluons, is consistent with the scenario that they have traversed 
through a dense matter that consists of colored quarks and gluons 
\cite{gyulassy1,levai1}. Conversions of minijet partons to high transverse 
momentum hadrons is usually modeled by fragmentation functions which 
describe how minijet partons combine with quarks and antiquarks from the 
vacuum to form hadrons as they separate. The parameters in the fragmentation 
functions can be determined by fitting the experimental data from high energy 
electron-positron annihilation and proton-proton collisions.  Because of 
the presence of the quark-gluon plasma, which is lacking in proton-proton 
collisions, minijet partons in heavy ion collisions can also combine with 
quarks and antiquarks in the QGP to form hadrons. Since the momenta of 
quarks and antiquarks in the QGP are much smaller than those 
of minijet partons, these hadrons have momenta between those from the 
independent fragmentations of minijet partons and from the hadronization 
of the QGP. These intermediate momentum hadrons thus carry information 
about the QGP formed in relativistic heavy ion collisions. 

In this letter, we shall adopt the coalescence model to study hadron 
production from the recombination of minijet partons with  
QGP partons. The coalescence model has previously been used in the ALCOR 
\cite{alcor} and MICOR \cite{micro} models to describe hadron yields 
from the QGP expected to be formed in relativistic heavy ion collisions. 
More recently, it has been introduced in a Multiphase Transport Model 
(AMPT) \cite{ampt} to model the hadronization of the partonic matter 
formed during the initial stage of relativistic heavy ion collisions 
\cite{melting}. Coalescence of hard minijet partons with soft QGP partons 
was first introduced in Ref. \cite{lin} to predict the flavor ordering of 
elliptic flows of high transverse momentum hadrons.  In the present 
study, we shall show that including hadrons from this production mechanism 
besides thermal hadrons from the hadronization of an expanding QGP can 
explain the increasing antiproton to pion ratio at low transverse
momentum and comparable antiproton and pion yields at intermediate 
transverse momentum observed in recent experimental data from RHIC 
\cite{ppi}. We further show that the antiproton to pion ratio decreases 
at large transverse momenta and approaches the small value given by 
the independent fragmentations of minijet partons.

\begin{figure}[ht]
\includegraphics[height=3in,width=2.5in,angle=-90]{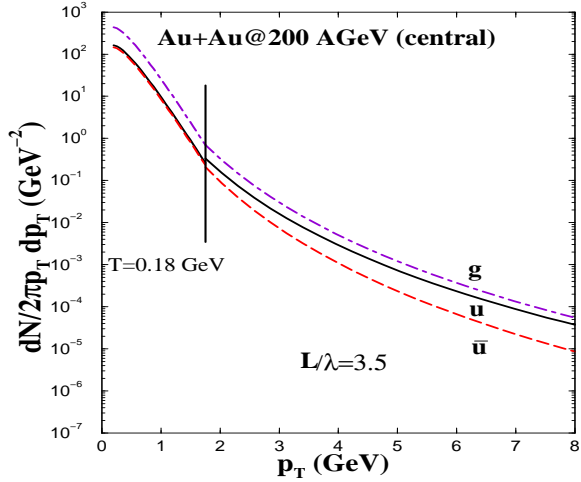}
\vspace{-0.3cm}
\caption{Parton transverse momentum distributions at hadronization
in Au+Au collisions at $\sqrt s=200$ AGeV. The minijet partons 
with $p_T$ greater than 1.75 GeV/c after energy loss 
are shown by dash-dotted curve for gluons, solid curve for $u$ quarks, 
and dashed curve for $\bar u$ quarks. Partons with $p_T$ 
below 1.75 GeV/c are from the quark-gluon plasma at temperature $T=180$ MeV.}
\label{parton}
%\vspace{-0.3cm}
\end{figure}

We consider central Au+Au collisions ($0-10\%$) at 200 AGeV. The 
transverse momentum distribution of minijet partons in midrapidity 
can be obtained from an improved perturbative QCD calculation \cite{yi02}
with a nucleon parton distribution function (PDF) that includes transverse 
momentum smearing. Kinematic details and a systematic analysis of $pp$ 
collisions can be found in Ref.~\cite{yi02}. Using the GRV94 LO result 
for the PDF \cite{structure} and the KKP fragmentation function from 
Ref.~\cite{KKP}, the measured data in the reaction $pp\to\pi^0X$ at 
$\sqrt{s}=200$ GeV can be reproduced with an average transverse momentum 
smearing of 1.4 GeV. To include parton energy loss in the QGP formed in 
the collisions, we use an effective opacity $L/\lambda=3.5$ as extracted 
from fitting the spectrum of high transverse momentum pions measured at 
RHIC \cite{levai1}. Taking the momentum cutoff $p_0=1.75$ GeV/c for the 
minijet partons, the transverse momentum spectra of minijet partons at 
midrapidity ($y=0$) in central Au+Au collisions at $\sqrt s=200$ AGeV are 
shown in Fig. \ref{parton} for the gluon (dash-dotted curve), $u$ 
(solid curve) and $\bar u$ (dashed curve) quarks. The $d$ and $\bar d$ 
quark distributions are similar to those of $u$ and $\bar u$ quarks. 

The QGP partons, which dominate transverse momentum below $p_0$, are 
taken to have a temperature $T=180$ MeV and to occupy a volume $V=730$ 
fm$^3$. This thermal parton transverse momentum distribution is also shown 
in Fig. \ref{parton}. Compared to the power-like minijet parton spectrum,
the spectrum of thermal partons is exponential. Because of scattering 
of minijet partons with thermal partons as they traverse the QGP, 
those with momenta around $p_0$ are expected to be thermalized with 
QGP partons, leading to a smooth spectrum around $p_0$. In the present 
schematic study, we neglect this effect. The total transverse energy per 
unit rapidity due to both thermal and minijet partons is about 570 GeV 
and is consistent with that measured in experiments \cite{transverse}. 
Most of this transverse energy comes from soft thermal partons as the 
contribution of minijet partons is only about 10\%.

The basic equation in the coalescence model for the formation of a
pion from a quark and an antiquark is similar to that for deuteron 
production from nucleons \cite{dover,baltz,mattiello} and can be written as 
\begin{eqnarray}
\frac{dN_\pi}{d^3{\bf p}_\pi}&=&g_\pi\int d^3{\bf x}_1d^3{\bf x}_2
d^3{\bf p}_1d^3{\bf p}_2f_q({\bf x}_1,{\bf p}_1)
f_{\bar q}({\bf x}_2,{\bf p}_2)\nonumber\\
&\times& \delta^3({\bf p}_\pi-{\bf p}_1-{\bf p}_2)
f_\pi({\bf x}_1-{\bf x}_2,{\bf p}_1-{\bf p}_2).
\end{eqnarray}
In the above, $f_q({\bf x},{\bf p})$ and $f_{\bar q}({\bf x},{\bf p})$ 
are the Wigner distribution functions for quarks and antiquarks, 
respectively, and they are normalized to their numbers, i.e.,  
$\int d^3{\bf x}d^3{\bf p}f_{q,\bar q}({\bf x},{\bf p})=N_{q,\bar q}$.
The Wigner function for the pion is denoted by $f_\pi({\bf x},{\bf p})$,
and it is normalized as  
$\int d^3{\bf x}d^3{\bf p}f_\pi({\bf x},{\bf p})=(2\pi)^3$.
The factor $g_\pi$ takes into account the probability of forming a
colorless spin zero pion from spin 1/2 color quarks, e.g., 
$g_{\pi^+}=1/36$ for forming a $\pi^+$ from a pair of $u$ and $\bar d$ 
quarks.

Assuming that quarks and antiquarks are uniformly distributed in the
fireball of volume $V$, then the quark and antiquark Wigner functions
are simply related to their momentum distributions, i.e., 
$f_{q,\bar q}({\bf x},{\bf p})=1/VdN_{q,\bar q}/d^3{\bf p}$.
The Wigner function for a pion depends on the spatial and momentum
distributions of its constituent quark and antiquark. For a schematic study,
we use uniform distributions, i.e., 
\begin{eqnarray}
f_\pi({\bf x},{\bf p})=\frac{9\pi}{2\Delta_x^3\Delta_p^3}
\Theta(\Delta_x-|{\bf x}|)\Theta(\Delta_p-|{\bf p}|),
\end{eqnarray}  
where $\Delta_x$ and $\Delta_p$ are, respectively, the spatial and
momentum cutoffs in the phase space of quark-antiquark relative 
motions. The uncertainty relation then requires $\Delta_x\Delta_p\ge\hbar$.

For hadrons in midrapidity ($y=0$), we only need to consider
their transverse momentum. Since minijet partons move essentially with
velocity of light, they are more likely to coalesce with 
comoving QGP partons, i.e., moving in the same direction, to form 
hadrons. This leads to the following simplified expression for coalescence 
of quarks and antiquarks to form pions:
\begin{eqnarray}
\frac{dN_\pi}{d^2{\bf p}_{\pi,T}}&=&g_\pi
\frac{6\pi^2}{V\Delta_p^3\,p_{\pi,T}}\int p_{q,T}dp_{q,T}p_{\bar q,T}
dp_{\bar q,T}\nonumber\\
&\times&\frac{dN_q}{d^2{\bf p}_{q,T}}\frac{dN_{\bar q}}{d^2{\bf p}_{\bar q,T}}
\delta(p_{\pi,T}-p_{q,T}-p_{\bar q,T})\nonumber\\
&\times&\Theta(\Delta_p-|p^\prime_{q,T}-p^\prime_{\bar q,T}|/2),
\end{eqnarray}
where $p^\prime_{q,T}$ and $p^\prime_{\bar q,T}$ are momenta of 
quark and antiquark in the center-of-mass frame of formed pion.
In making the Lorentz transformation from the fireball frame to the
pion frame, we use the current quark mass of 10 MeV for minijet quarks
but the constituent quark mass of 300 MeV for QGP quarks due to 
nonperturbative effects. 

The above results can be generalized to formation of protons and
antiprotons from the parton distribution functions. We take the antiproton
Wigner function as
\begin{eqnarray}
f_{\bar p}({\bf x},{\bf y};{\bf p},{\bf q})&=&\frac{9\pi}
{2\Delta_x^3\Delta_p^3}\Theta(\Delta_x-|{\bf x}|)
\Theta(\Delta_p-|{\bf p}|)\nonumber\\
&\times& \frac{9\pi}{2\Delta_y^3\Delta_q^3}\Theta(\Delta_y-|{\bf y}|)
\Theta(\Delta_q-|{\bf q}|),
\end{eqnarray}  
where ${\bf x}=({\bf x_1}-{\bf x_2})$ and 
${\bf y}=({\bf x_1}+{\bf x_2})/2-{\bf x_3}$ are the relative coordinates 
among three antiquarks, while 
${\bf p}=({\bf p_1}-{\bf p_2})/2$ and 
${\bf q}=({\bf p_1}+{\bf p_2}-{\bf p_3})/2$ are corresponding 
relative momenta. As for pions, the uncertainty relation requires that
$\Delta_x\Delta_p$ and $\Delta_y\Delta_q$ cannot be less than
$\hbar$. For simplicity, we take $\Delta_x=\Delta_y$ and 
$\Delta_p=\Delta_q$. This leads to the following antiproton transverse 
momentum spectrum from coalescence of three antiquarks:
\begin{eqnarray}
\frac{dN_{\bar p}}{d^2{\bf p}_{\bar p,T}}&=&g_{\bar p}
\frac{36\pi^4}{V^2\Delta_p^6\,p^2_{\bar p,T}}
\int \prod_{i=1,3}p_{\bar q_i,T}dp_{\bar q_i,T}\frac{dN_{\bar q_i}}
{d^2{\bf p}_{\bar q_i,T}}\nonumber\\
&\times&\delta(p_{\bar p,T}-p_{\bar q_1,T}-p_{\bar q_2,T}-p_{\bar q_3,T})
\nonumber\\
&\times&\Theta(\Delta_p-|p^\prime_{\bar q_1,T}-p^\prime_{\bar q_2,T}|/2)
\nonumber\\
&\times&\Theta(\Delta_p-|p^\prime_{\bar q_1,T}+p^\prime_{\bar q_2,T}
-p^\prime_{\bar q_3,T}|/2).
\end{eqnarray}
In the above, $g_{\bar p}$ is the probability of forming a colorless 
spin 1/2 antiproton from two $\bar u$ quarks and one $\bar d$ quark,
i.e., $g_{\bar p}=1/108$.  The antiquark momenta in the center-of-mass 
frame of formed antiproton are denoted by $p^\prime_{\bar q_i,T}$.

\begin{figure}[ht]
\includegraphics[height=3in,width=2.5in,angle=-90]{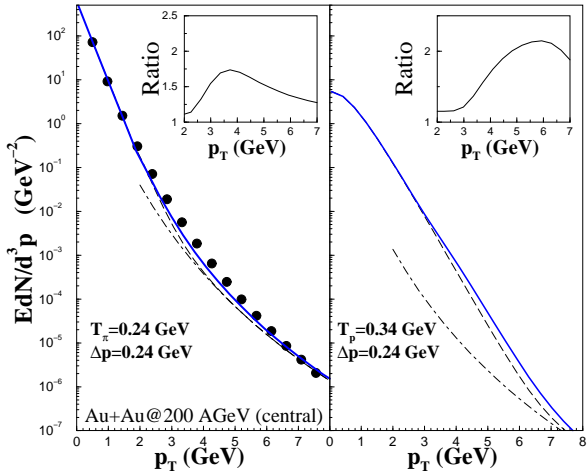}
\vspace{-0.3cm}
\caption{Transverse momentum spectra of pions (left panel)
and antiprotons (right panel) from Au+Au collisions at 
$\sqrt s=200$ AGeV. Dashed curves are results including contributions
from thermal hadrons and hadrons from independent fragmentations of minijet 
partons (dash-dotted curves). Adding also hadrons from coalescence of minijet 
partons with QGP partons gives the solid curves. Ratio of the solid to
the dashed curve is given in the insets. Experimental $\pi^0$ data
\cite{phenix1} are shown by filled circles.}
\label{hadron}
%\vspace{-0.3cm}
\end{figure}

We first take thermal hadrons to have exponential transverse mass 
($m_T=\sqrt{m^2+p_T^2}$) spectra with inverse slope parameter equal to 
240 MeV for pions and 340 MeV for antiprotons in order to reproduce the 
observed low transverse momentum spectra of these particles. The larger 
inverse slope parameter than the QGP temperature reflects the effect due 
to collective transverse flow of the QGP, which increases the transverse 
momentum of heavier antiprotons more than that of lighter pions. The 
number of thermal pions is fixed by fitting the measured pion spectrum 
at low transverse momenta, while that of thermal antiprotons is determined 
by requiring that it is 0.7 of the pion number at transverse momentum of 
2 GeV/c as in experimental measurements.  Hard hadrons, shown in 
Fig. \ref{hadron} by dash-dotted curves, are obtained from minijet 
partons using the KKP fragmentation function of Ref.~\cite{KKP}, which 
has been shown to reproduce measured high transverse momentum particles 
at RHIC. The hadron spectra from the sum of these two contributions are 
shown in Fig. \ref{hadron} by dashed curves. As shown in the left panel, 
the final pion spectrum at intermediate transverse momentum is below the 
experimental data from the PHENIX Collaboration \cite{phenix1} shown by 
filled circles. 

To evaluate the contribution to pion and antiproton production from 
coalescence of minijet quarks and antiquarks with those from the QGP, 
the effect due to gluons is taken into account by converting them to 
quark-antiquark pairs with equal probability for different flavors as 
assumed in the ALCOR model \cite{alcor}. Using $\Delta_p=0.24$ GeV/c 
corresponding to a pion size of $\Delta_x\approx 0.85$ fm, the coalescence 
contribution leads to a final pion spectrum, shown by the solid curve 
in the left panel of Fig. \ref{hadron}, that agrees reasonably with 
the experimental data. For simplicity, the same $\Delta_p$ is used to 
evaluate the coalescence contribution to antiproton production, and 
the resulting final antiproton spectrum is shown by the solid curve in 
the right panel of Fig. \ref{hadron}. From the ratio of hadron spectra 
with and without the coalescence contribution shown in the insets of 
Fig. \ref{hadron}, this new mechanism is seen to enhance the yields of 
intermediate transverse momentum pions and antiprotons by factors of 
about 1.7 and 2.2, respectively.

\begin{figure}[ht]
\includegraphics[height=3in,width=2.5in,angle=-90]{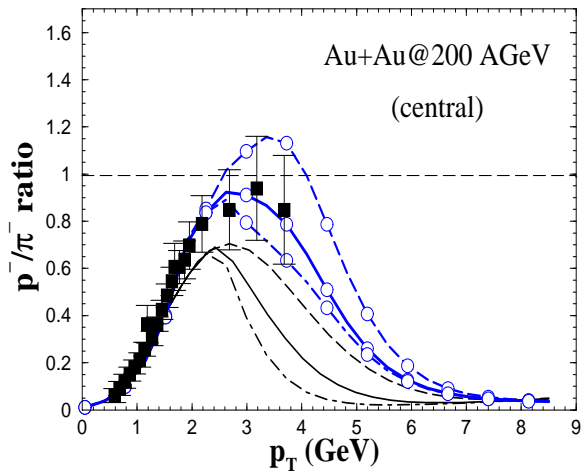}
\vspace{-0.3cm}
\caption{Antiproton to pion ratios from Au+Au collisions at $\sqrt s=200$ 
AGeV. Dash-dotted, solid, and dashed curves are results using, respectively,
240, 300, and 340 MeV for the inverse slope parameter of intermediate 
$p_T$ antiprotons. Corresponding results including also
contributions from coalescence of minijet and QGP partons are shown
with open circles. Filled squares are the experimental data \cite{ppi}.}
\label{ratio}
%\vspace{-0.3cm}
\end{figure}

In Fig. \ref{ratio}, we show the ratio of antiproton to pion as a
function of transverse momentum obtained with different scenarios for
the antiproton spectrum but the same empirical pion spectrum shown in
the left panel of Fig. \ref{hadron}. The dashed curve corresponds
to an antiproton spectrum that includes only the thermal and perturbative 
ones, i.e., the dashed curve in the right panel of Fig. \ref{hadron}. 
This theoretical result is below the experimental antiproton to pion 
ratio given by filled squares \cite{ppi} when the transverse momentum 
is above 2 GeV/c. Including antiproton production from coalescence 
of minijet partons with thermal partons enhances the ratio significantly 
as shown by the dashed curve with open circles, which is now above the 
experimental ratio.

Since the collective flow effect is expected to diminish with increasing 
antiproton transverse momentum, we explore this possibility by varying the 
inverse slope parameter of the transverse momentum spectrum of thermal
antiprotons above 2.5 GeV/c. Taking the inverse slope parameter of these 
antiprotons to be 240 MeV, same as that for pions, the antiproton to pion 
ratio is shown in Fig. \ref{ratio} by the dash-dotted curve and the  
dash-dotted curve with open circle for the cases without and with the 
coalescence contribution to antiproton production, respectively. In this 
case, coalescence of minijet and QGP partons again increases appreciably
the antiproton to pion ratio at intermediate transverse momenta compared 
to that without this contribution, bringing the theoretical results
closer to the experimental data. If the inverse slope parameter of 
intermediate transverse momentum thermal antiprotons is taken to 
have a value of 300 MeV, between those of low transverse momentum thermal 
pions and antiprotons, the resulting antiproton to pion ratio is
shown by the solid curve and the solid curve with open circles in 
Fig. \ref{ratio} for the cases  without and with the coalescence 
contribution. It is seen that the experimental data can now be reproduced 
with the inclusion of antiproton production from coalescence of minijet 
partons with thermal partons.

The antiproton to pion anomaly has previously been attributed to enhanced 
production of antiprotons with intermediate transverse momentum from the 
baryon junctions in incident nucleons \cite{vitev}. The possibility of 
enhanced baryon to pion ratio due to parton coalescence was suggested 
in Ref. \cite{voloshin}. Using a parton distribution function that is 
fitted to measured pion transverse momentum spectrum, a parton 
recombination model similar to the coalescence model indeed leads 
to a large antiproton to pion ratio at intermediate transverse momenta
\cite{hwa}. In Ref. \cite{fries}, the antiproton to pion anomaly is
explained by the recombination of only QGP partons with a high effective
temperature. Further studies are needed to find out which of these 
mechanisms including ours based on the coalescence of minijet partons
with partons from the qaurk-gluon plasma are most relevant to  
the observed large anitproton to pion ratio.

In summary, we have proposed a mechanism for the hadronization
of minijet partons in heavy ion collisions at relativistic energies. 
Instead of usual independent fragmentations, in which they combine
with quarks and antiquarks from the vacuum to form hadrons, these minijet
partons are allowed to coalescence with thermal quarks and antiquarks 
from the quark-gluon plasma created in the collisions to form hadrons.
Using the minijet partons predicted from the perturbative QCD, we find
that this mechanism is important for production of hadrons with
intermediate transverse momentum, leading to comparable antiproton and 
pion yields in this momentum region as observed experimentally. It further 
predicts that the antiproton to pion ratio would decrease as their transverse 
momenta become large. In this high transverse momentum region,
independent fragmentations of minijet partons dominate particle
production and lead to a very small antiproton to pion ratio. Confirmation 
of this hadronization mechanism thus provides another evidence for the 
formation of the quark-gluon plasma in relativistic heavy ion collisions.

%\bigskip

We are grateful to Su Houng Lee for useful discussions. P.L. would 
also like to thank discussions with G. Papp and G. Fai, and the warm 
hospitality of the Cyclotron Institute at Texas A\&M University during 
his visit.  This paper is based on work supported in part by the U.S. 
National Science Foundation under Grant No. PHY-0098805 and the Welch 
Foundation under Grant No. A-1358. The work of V.G. is further supported 
by a fellowship from the National Institute of Nuclear Physics (INFN) 
in Italy, while that of P.L. by Hungarian OTKA Grant Nos. T034269 
and T043455.

\end{document}